\documentclass[11pt, oneside]{amsart}   	
\usepackage{geometry}                		
\usepackage{graphicx}				
\usepackage{amssymb,amsmath,url}

\title{Comparative Analysis of Dengue versus Chikungunya 
Outbreaks in Costa Rica
}

\author{Fabio S\'{a}nchez}  
\address{ Escuela de Matematica \\
		Universidad de Costa Rica\\
		Ciudad Universitaria Rodrigo Facio\\
		Codigo Postal 2060\\
		San Jose, Costa Rica\\
		America Central \\	
              Tel.: +506-2511-6608\\
}
 \email{fabio.sanchez@ucr.ac.cr} 
      
\author{       Luis A. Barboza } 
\address{Escuela de Matematica \\
		Universidad de Costa Rica\\
		Ciudad Universitaria Rodrigo Facio\\
		Codigo Postal 2060\\
		San Jose, Costa Rica\\
		America Central 	
		}

\author{        David Burton }
\address{Department of Mathematics and Statistics\\
		Box 70663\\
		East Tennessee State University\\
		Johnson City, TN 37614-0663        
		}
\thanks{D. Burton was partially funded by National Science Foundation grant number DUE-1356397.}		
\author{        Ariel Cintr\'{o}n-Arias}
\address{Department of Mathematics and Statistics\\
		Box 70663\\
		East Tennessee State University\\
		Johnson City, TN 37614-0663        
		}


\begin{document}
\maketitle

\begin{abstract}
For decades, dengue virus has been a cause of major public health concern in 
Costa Rica, due to its landscape and climatic conditions that favor the 
circumstances in which the vector, {\it Aedes aegypti}, thrives. The emergence 
and introduction throughout tropical and subtropical countries of the 
chikungunya virus, as of 2014, challenged Costa Rican health authorities to 
provide a correct diagnosis since it is also transmitted by the same 
vector and infected hosts may share similar symptoms. We study the 
2015-2016 dengue and chikungunya outbreaks in Costa Rica while 
establishing how point estimates of  
epidemic parameters for both diseases compare to one another. Longitudinal 
weekly incidence reports of these outbreaks signal likely misdiagnosis of 
infected individuals: underreporting of chikungunya cases, while overreporting
cases of dengue. Our comparative analysis is formulated with a 
single-outbreak deterministic model that features an {\it undiagnosed} class. 
Additionally, we also used a {\it genetic algorithm} in the context of weighted 
least squares to calculate point estimates of key model parameters and initial 
conditions, while formally quantifying misdiagnosis.
\end{abstract}

\section{Introduction}\label{intro}
Dengue and chikungunya are vector-borne diseases transmitted primarily by the mosquito {\it Aedes (Ae.) aegypti}, which has successfully invaded the vast majority of countries in the tropics and sub-tropics~\cite{cdcdengue,harris2000}. Dengue affects approximately 100 million people annually~\cite{cdcdengue}. The distribution of reported dengue and chikungunya cases has spread into countries with geographical features fostering favorable mosquito habitats like Costa Rica \cite{mir2014}. The chikungunya virus since it was introduced in the Americas in late 2013 has spread to forty five countries with more than 1.7 million unconfirmed cases~\cite{cdcchikv}.

Costa Rica is located in Central America, it has a population of approximately 4.9 million and an area of $51,100$ sq. km~\cite{mscr}. The weather is tropical, which makes it ideal for the vectors ({\it Ae. aegypti} and {\it Ae. albopictus}) that transmit dengue and chikungunya virus. Dengue fever has reshaped public health policy in the country since it first emerged more than two decades ago~\cite{cdcdengue}. The implementation of prevention and educational policies has increased over the years. Awareness of the disease is ample in most endemic areas of the country~\cite{mscr}.

More recently, another vector-borne disease has spread throughout the Americas. In 2014 the chikungunya virus, which is also transmitted by the {\it Ae. aegypti} mosquito, was introduced in Costa Rica. The disease is spreading rapidly and poses substantial challenges to health officials when conducting diagnosis. Costa Rica reported $145$ cases in 2014, compared to $4,912$ reported cases in $2015$~\cite{mscr}.

In $2015$ the chikungunya virus followed a similar trend to previous dengue outbreaks~\cite{mscr}. This is particularly important because the newly introduced virus, chikungunya, could be misdiagnosed as dengue due to the similarity of symptoms~\cite{cdcchikv,cdcdengue}. This is critical for public health officials since correctly diagnosing the disease is vital for proper treatment. There are also studies that suggest that co-infection is possible between the viruses~\cite{chahar2009,hapuarachchi2008}. 

In this paper, we study and compare the {\it dengue} and {\it chikungunya} reported cases during 2015-2016 in 
Costa Rica. We explore the importance of {\it undiagnosed} cases, in part because the majority of cases are suspected of being 
asymptomatic or individuals who may show mild symptoms and do not go to a hospital for treatment~\cite{cdcdengue}. We use a 
single outbreak deterministic model and fit the model with incidence data. We also compute point estimates of
key parameters and initial conditions, while implementing global and local optimization algorithms within the context of a weighted least squares scheme.

\section{Mathematical Model}\label{model}

\begin{figure}[h]
\centering
\includegraphics[width=3in,height=2in]{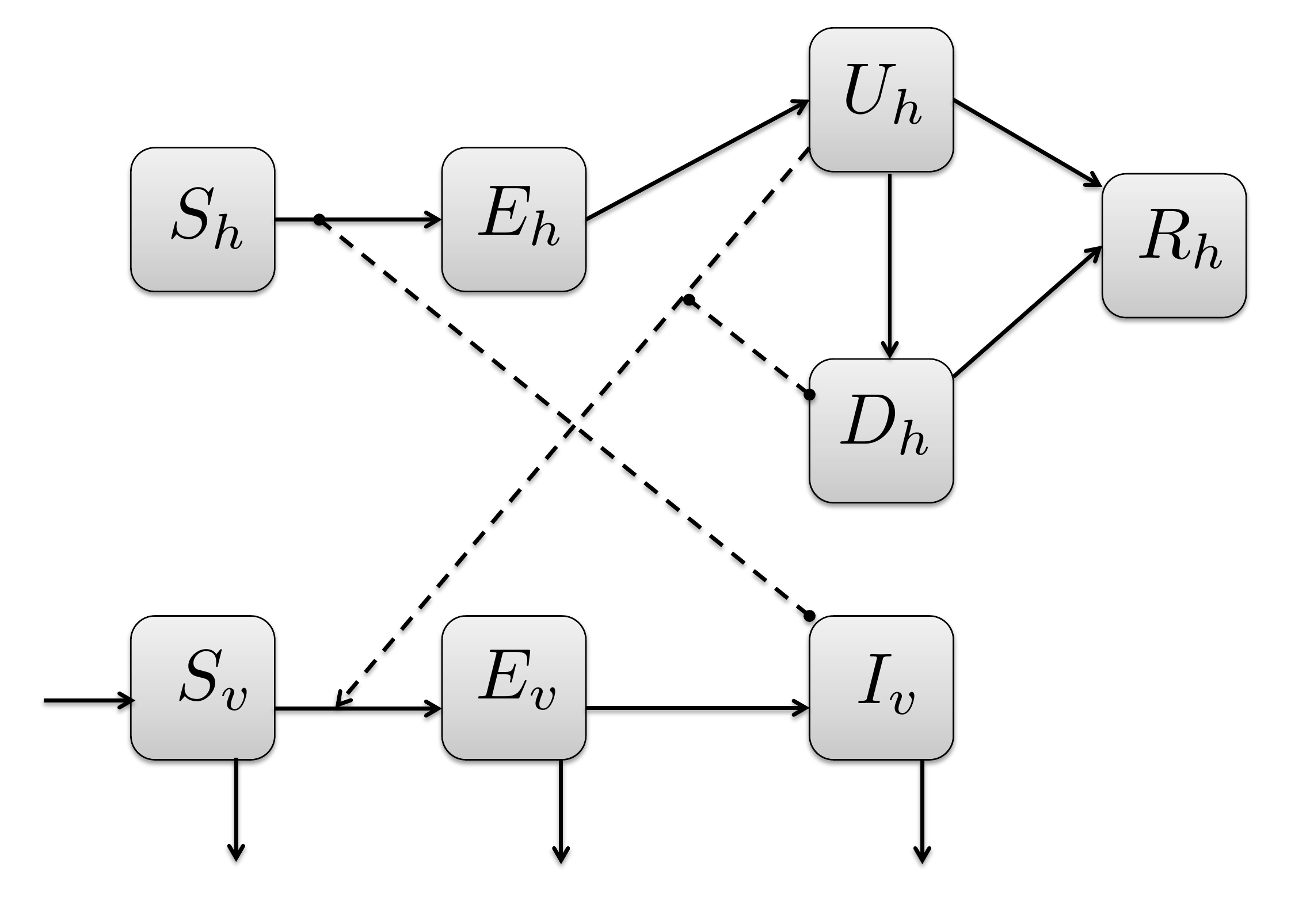}
\caption{Compartments of a deterministic single-outbreak model for
a host-vector system.  State variables are defined in text.}
\label{compdiag}      
\end{figure}

We introduce a single-outbreak deterministic model to fit data from the 2015-2016 dengue and chikungunya outbreaks in Costa Rica. There are five classes 
that describe the host dynamics: $S_h$, susceptible hosts, $E_h$, exposed hosts, $U_h$, undiagnosed hosts, $D_h$, diagnosed hosts, and $R_h$, 
recovered hosts. Here, we assume that one can only be infected by one strain of the dengue virus. The state variables describing the vector 
dynamics are: $S_v$, susceptible vectors, $E_v$, exposed/latent vectors, and $I_v$, infected vectors.

The model assumes homogeneous mixing, in other words, encounters among members of all classes are equally likely to occur without 
any preference. Such encounters without preference are in a sense equivalent to simple random sampling. The rate of change of 
each class with respect to time $t$ is denoted by $d/dt$. The following system of nonlinear differential equations (rates of change) define 
the single-outbreak model that we propose here:
\begin{eqnarray} \label{sh}
\frac{S_h}{dt} &=& -\beta S_h \frac{I_v}{N_v}, \\
\frac{E_h}{dt} &=& \beta S_h \frac{I_v}{N_v}-\alpha_h E_h, \\
\frac{U_h}{dt} &=& \alpha_h E_h-(\gamma+\delta) U_h, \\
\frac{D_h}{dt} &=& \delta U_h-\gamma D_h,\\
\frac{R_h}{dt} &=& \gamma U_h+\gamma D_h, \\
\frac{S_v}{dt} &=& \mu N_v-\beta_v S_v \frac{(U_h+D_h)}{N_h}-\mu S_v, \\
\frac{E_v}{dt} &=& \beta S_v \frac{(U_h+D_h)}{N_h}-(\mu+\alpha_v) E_v, \\
\label{iv}
\frac{I_v}{dt} &=& \alpha_v E_v-\mu I_v,
\end{eqnarray}
where $N_h=S_h+E_h+U_h+D_h+R_h$ and $N_v=S_v+E_v+I_v$.  Previous dengue 
models focus on a more theoretical 
approach, including but limiting to, linear stability analysis, 
bifurcation analysis, next generation operators, etc, 
\cite{esteva1998,pawelek,murillo2014,sanchez2006,fsanchez2012}.  Additional
studies that factor in climatological 
variables are \cite{chowell2006b,chowell2004,chowell2006}.

Because of the time scale typical of an outbreak, of nearly a year, we assume no births or deaths take place in the host population, hence, 
the population remains constant throughout the outbreak. On other hand, the vector population is also assumed to be constant, however, 
the number of births is set to match the number of deaths, thus reflecting the fast dynamics of the vector relative to the host population dynamics.

Transmission dynamics is assumed to be the same for both viruses, dengue and chikungunya. Hosts who are susceptible ($S_h$) can 
obtain the virus from an infected vector ($I_v$). Infection is considered here with two stages: latency ($E_h$ and $E_v$,), where hosts 
and vectors are unable to transmit the virus; and infectiousness, where the virus can be transmitted. Infectious vectors are part of 
the $I_v$ class. For the host population the infectious class is divided into: {\it undiagnosed} ($U_h$) and {\it diagnosed} ($D_h$). After going 
through these latter stages the host recovers ($R_h$). We assume permanent immunity for both viruses (dengue and chikungunya) 
and do not allow for reinfection. 

\section{Data and Methodology}
\begin{figure}[htb!]
\centering
\includegraphics[width=3.5in,height=2.5in]{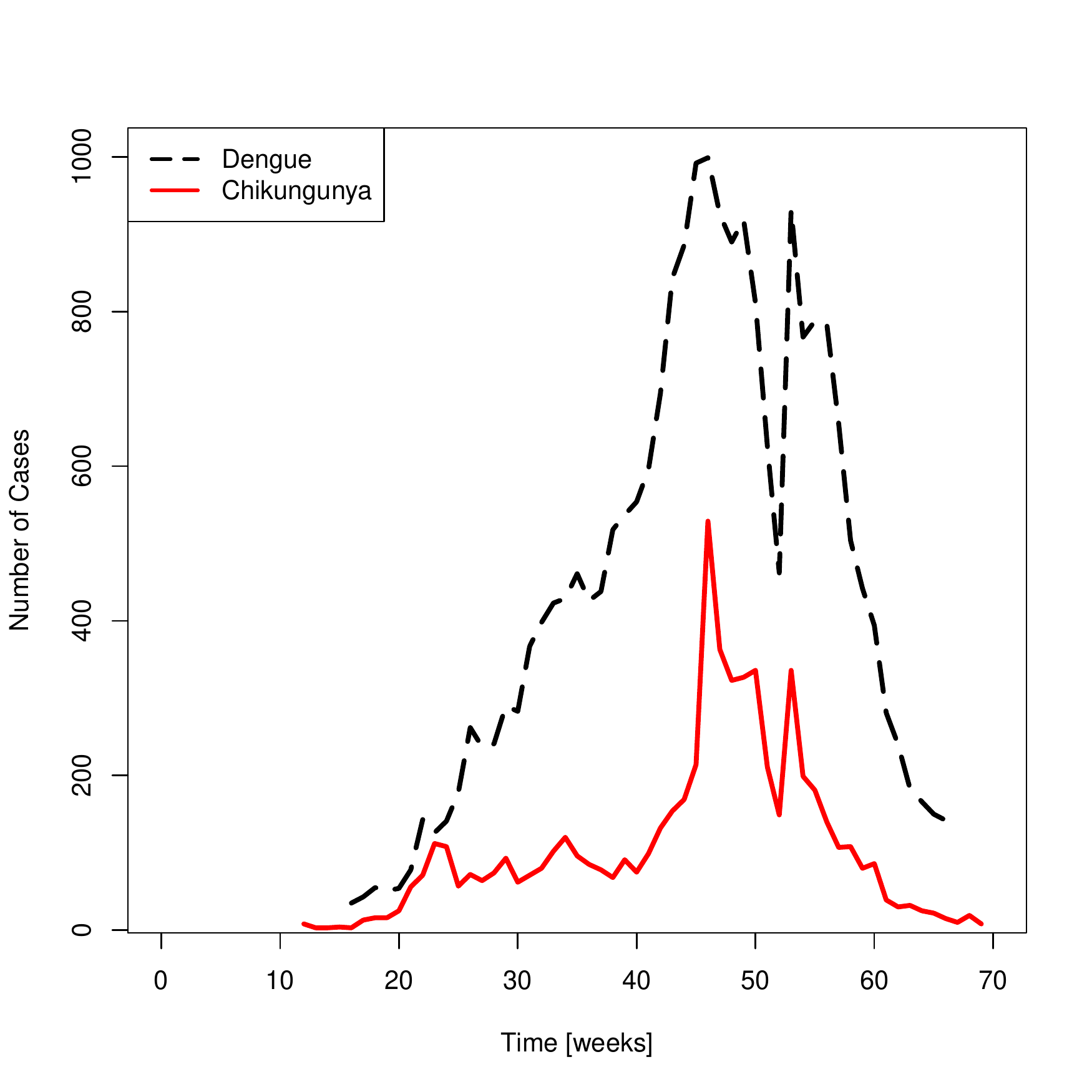}
\caption{Cases of dengue and chikungunya reported weekly during 2015-2016~\cite{mscr}. }
\label{data}       
\end{figure}

The number of weekly reported counts (of laboratory and clinically diagnosed cases) of dengue and chikungunya during the 2015-2016 outbreak in Costa Rica were employed. These observations were reported from May 2015 through May 2016 by Costa Rica's Department of Health and Human Services (``Ministerio de Salud de Costa Rica")~\cite{mscr}. Figure~\ref{data} displays the number of cases versus time. In its left panel, cases of dengue appear depicted in a dashed curve, while cases of chikungunya are displayed in a solid curve. The time unit in Figure~\ref{data} is weeks, labeled in the chronological order of the calendar year. Thus, observations in the time interval from week 10 to week 52 correspond to measurements from 2015, while data points after week 52 denote counts from 2016.
Dengue is seasonal and starts roughly when the rainy season begins (around week 16 of the calendar year). Although climate variability in the country is an important factor we do not address it in this study.  Instead,
we mostly consider the incidence data \lq\lq anomalies\rq\rq (in the context of misdiagnosis, i.e., measurement error) of the aforementioned outbreak.

The onset of dengue and chikungunya symptoms resembles those of a simple cold or a mild to moderate flu infection. For example, headache, fatigue, and fever are common symptoms. Only when symptoms become severe it is then more reliable to distinguish between the two viruses. 

\begin{figure}[htb!]
\centering
\includegraphics[width=2in,height=2in]{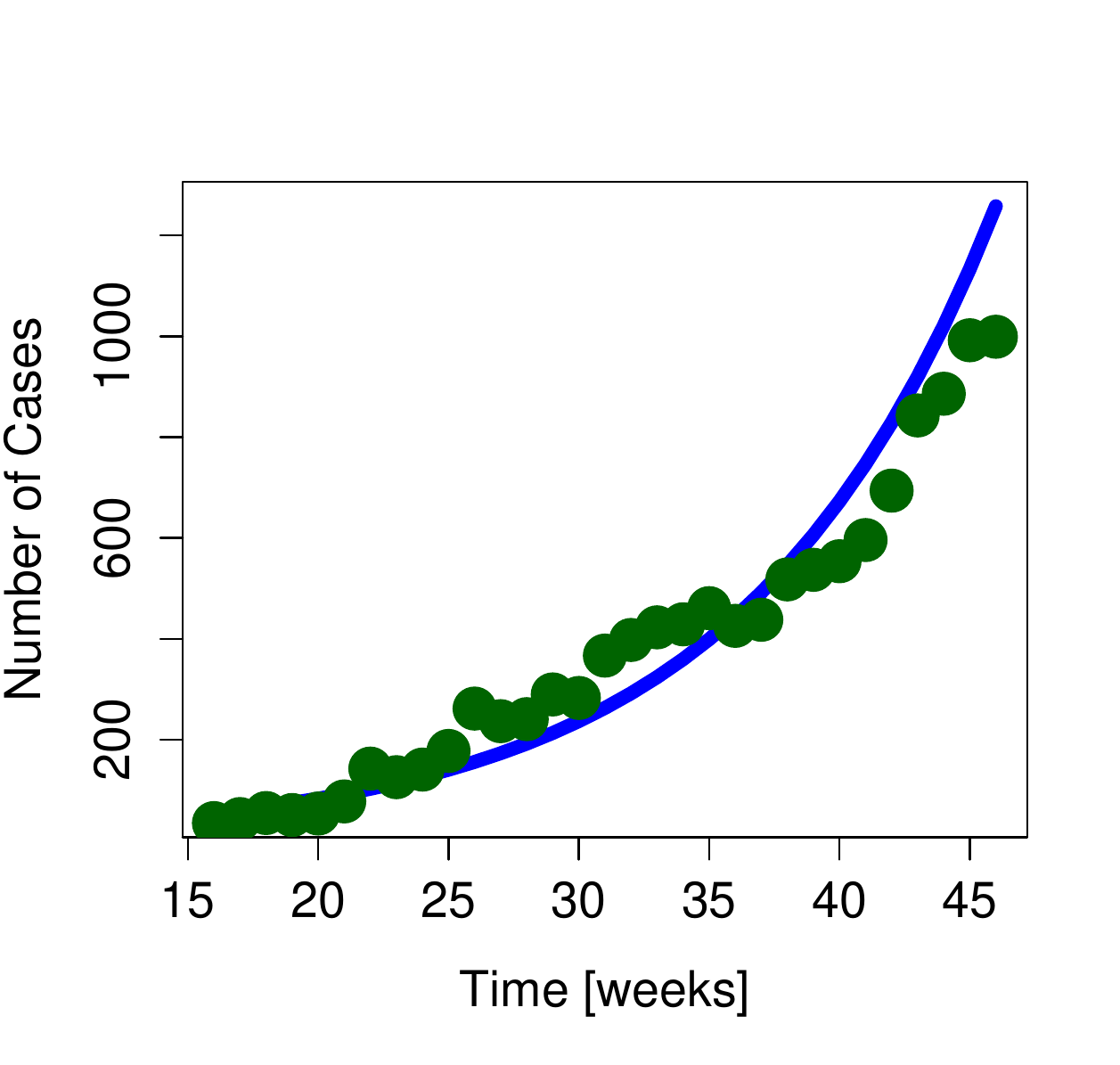}
\includegraphics[width=2in,height=2in]{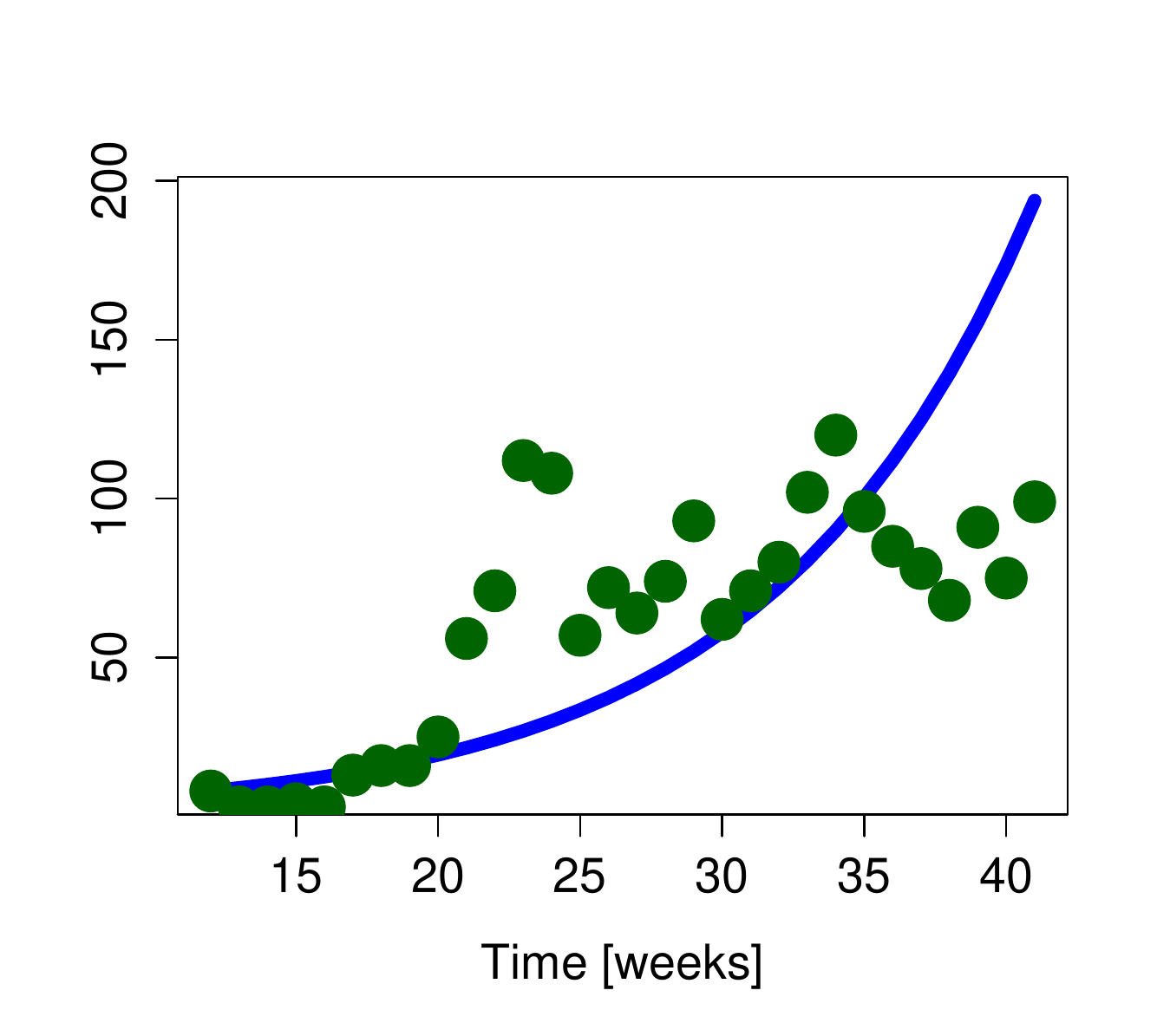}
\caption{Initial growth of the outbreaks in the first half of the 2015-2016 season. Left panel: Exponential growth of the dengue outbreak from calendar week 16 through 46. Right panel: initial growth of the chikungunya outbreak from calendar week 12 through 41.}
\label{initial_epi_growth}      
\end{figure}




The first well-known sign of an epidemic process during its initial stage is
exponential growth.  Despite the fact that Costa Rica has seen a number of large 
outbreaks in the last few years, during 2015 dengue cases display an 
abnormally high initial growth~\cite{mscr}.  The left panel of
Figure~\ref{initial_epi_growth} displays 
$x(t) = 10.3779 e^{ 0.1043t}$ versus $t$ together with the longitudinal observations
of dengue cases, where it is confirmed that the
initial growth obeys an exponential law.  The calculation of this exponential
function was derived from a linear regression model fitted to the incidence 
data in logarithmic scale, which had a value of the 
statistic $R^2 = 0.9294$.  Thus, implying a linear model was remarkably 
well-suited to
describe the number of cases in logarithmic scale, and then confirming
the appropriate exponentially growing trend (of the incidence data) in regular 
scale.  On the other hand, 
for chikungunya, the data does not follow exponential growth.  The right panel
of Figure~\ref{initial_epi_growth} depicts longitudinal reported cases
of chikungunya (circules) together with an exponential function (solid curve) that
was also derived from a best fit regression line.  Visual inspection
easily confirms what the value of the statistic $R^2 = 0.6365$ quantifies:
longitudinal cases of chikungunya are not appropriately described by exponentially
growing trends.  In summary, we have two scenarios: one overly optimistic case
count for dengue with $R^2 = 0.9294$, likely due to overreporting; and another
one where case counts are poorly exponentially growing with $R^2 = 0.6365$, failing to portray the most fundamental features of infectious diseases, but likely due to under reporting.

Additionally, having $R^2 = 0.6365$ is the first signal of an anomaly
with chikungunya cases in Costa Rica during the 2015-2016 outbreak.  In 
fact, this raises the question about the possibility of misdiagnosis 
between the two viruses transmitted by the same vector ({\it Ae. aegypti}). Perhaps even more interesting is the behaviour of the reported cases of 
chikungunya. Between weeks 25-35, data shows relatively steady number of 
cases (see solid curve in Figure~\ref{data}). The fact that chikungunya was recently 
introduced would suggest the population is mostly susceptible. However, 
early on 2015 we see that the disease takes off after week 35 and follows a similar trend as dengue with fewer cases being reported.

It is important to note that Costa Rican public health officials previously diagnosed dengue as the only disease transmitted by {\it Ae. aegypti} until late 2014. It is possible that some if not many cases were diagnosed as
dengue instead of chikungunya due to the lack of familiarity of the recently introduced disease.

The vector $\theta = (\beta,\delta,\mu,S_{h}(0),S_{v}(0))$ stores the unknown model parameters and initial conditions of the dynamical system for dengue disease. In the case of chikungunya, the parameter vector is defined as $\theta = (\beta,\delta,S_{h}(0),S_{v}(0))$. To estimate them, from longitudinal data, we consider a weighted least squares scheme where the incidence measurements at time points $t_k$ (for $k=1,\dots,n$) define a random observation process, denoted as $Y_k$. To take in account the measurement error and due to the integer-valued nature of the process, we assume that $Y_k$ behaves as a Poisson random variable where its mean is obtained as result of model output. More specifically, 
	\begin{equation} \label{ols}
		Y_k \sim \text{Poisson}\left[z(t_k, \theta_0)\right],
	\end{equation}
	where $z(t_k, \theta)$ is the expected number of cases from $t=t_{k-1}$ to $t=t_k$, which is defined as
	\begin{equation} \label{cases}
		z(t_k, \theta) =   \int^{t_{k}}_{t_{k-1}} \delta \, U_h(t,\theta) dt. 
	\end{equation}
The observation process in equation (\ref{ols}) entails the evaluation of the model output at the true parameters, that is, $\theta = \theta_0$. A realization of the observation process, $Y_{k}=y_{k}$, can be employed to solve the box-constrained minimization problem,
	\begin{equation} \label{sumsq}
		\min_{\theta \in \mathcal{S}} \sum_{k=1}^{n} \frac{\left(y_k-z(t_k, \theta)\right)^2}{z(t_k, \theta)}.
	\end{equation}
A solution of (\ref{sumsq}), referred to as $\hat \theta$, is an estimate and serves as an approximation of the the true parameter $\theta_0$. Several authors have used similar procedures to estimate the parameters of dynamical systems for other diseases, see for example~\cite{banks2014,chowell2006b,chowell2004,pawelek}, among others. We used two successive methods of optimization in order to obtain the estimate $\hat \theta$: Genetic Algorithm~\cite{Holland92,scrucca} and the quasi-Newton method L-BFGS-B~\cite{byrd95}. The first algorithm provides an initial search on the parametric space, in order to skip possible local optima. The second algorithm takes as initial value the one obtained using GA and seeks to improve the search of the overall optimum.

\section{Results} \label{results}
We analyzed the possible misdiagnosis in dengue and chikungunya weekly reported data using a single-outbreak deterministic model for the 2015-2016 outbreak in Costa Rica. Using a genetic algorithm and weighted ordinary least squares we fit dengue and chikungunya reported cases to compute point estimates of key 
model parameters and highlight the importance of initial conditions for the diseases to thrive in a mostly susceptible population.  We do not report uncertainty quantification, i.e., neither confidence
intervals nor uncertainty bounds are provided for each point estimate.

\begin{figure*}[htb!]
\centering
\includegraphics[height=4in,width=4.5in]{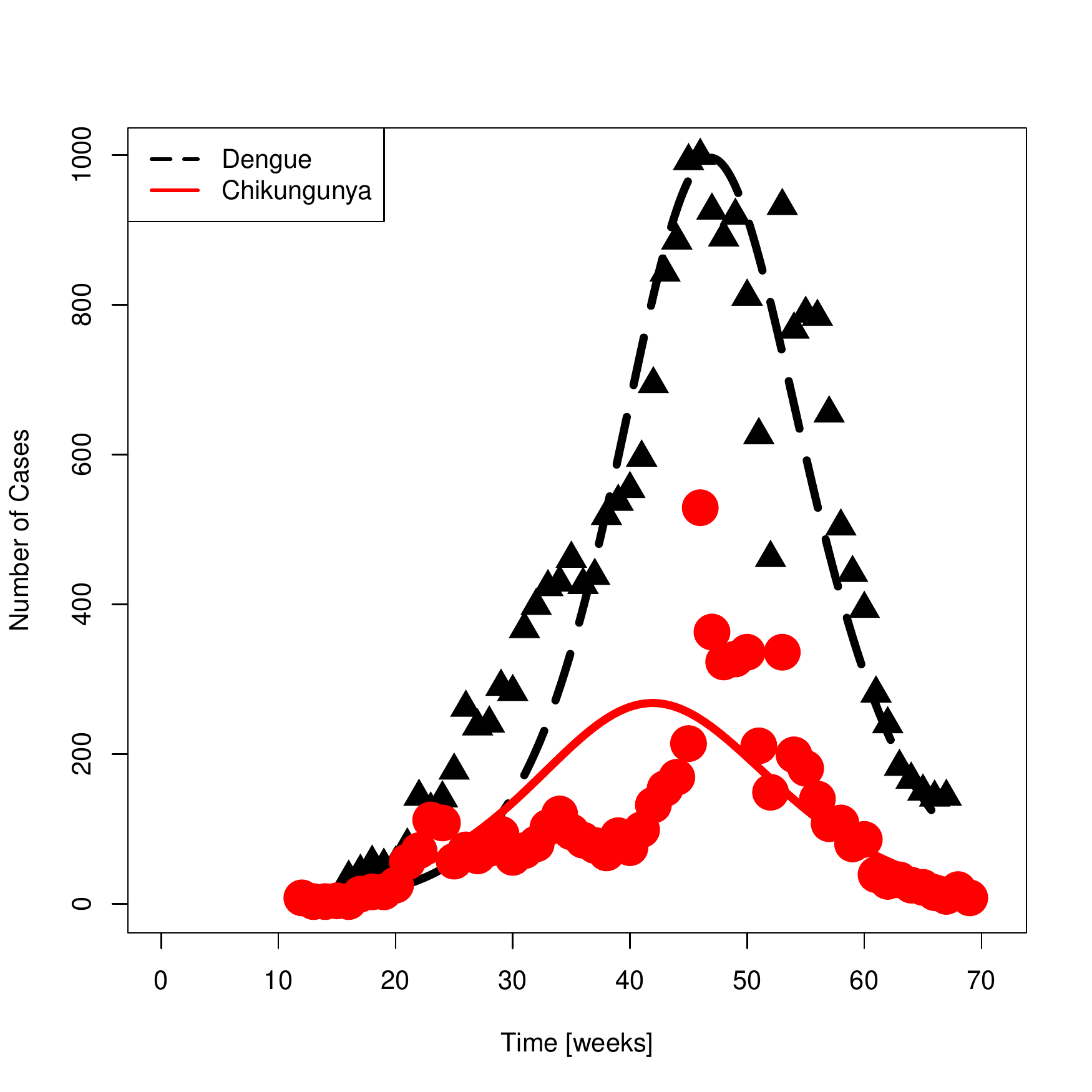}
\caption{Best fit curves or predicted cases, obtained by a
{\it genetic algorithm} in the context of weighted least squares 
optimization, appear as dashed and 
solid curves for dengue and chikungunya, respectively.  Reported
cases are depicted as triangules and circles, respectively.}
\label{fig:test}
\end{figure*}

Figure \ref{fig:test} depicts longitudinal observations and best fit curves for 
dengue and chikungunya.
Between calendar weeks 20 and 30 the model prediction is lower than the reported 
cases of dengue. This coincides with the stagnation of chikungunya cases as seen in Figure~\ref{data} during the same time period. The point 
estimate of the transmission rate, $\beta=4.745$, is within the observed range in 
previous studies \cite{manore2014}. The initial number of susceptible hosts, $S_h(0)$, and  
susceptible vectors, $S_v(0)$ were also estimated and appear reported in Table~\ref{partable}. 

Furthermore, the rate at which undiagnosed individuals become diagnosed, $\delta$ was estimated. The point estimate this parameter suggests that individuals are being diagnosed at a very fast rate for dengue, ($\delta= 275.979$). Although, it is known that most dengue cases are asymptomatic and therefore go undiagnosed~\cite{chastel2012}, this point estimate suggests that data anomalies are likely and dengue cases were over reported in the 2015-2016 outbreak. For chikungunya the estimated rate of diagnosis, $\delta=20.000$,
was much lower and hence, less individuals were diagnosed .

\begin{table}
	\centering
	\caption{Model parameters and initial conditions.  Units are in weeks.
	Known values are used to fix some parameters and initial conditions.
	Unknown quantities are considered active parameters in the optimization.
	Lower and upper bounds of epidemiological parameters are set in accordance to~\cite{manore2014}, they are employed here to define feasible ranges. Point estimates only, without uncertainty quantification, are reported here. N/A stands for ``it does not apply".}
	\label{partable}
	\begin{tabular}{|c|c|c|r|} \hline
		Parameter & Fixed Value & Range & Point Estimate \\ \hline
		\multicolumn{4}{|c|}{\textbf{Dengue}} \\ \hline\hline
		\multicolumn{4}{|l|}{Fixed Parameters} \\ \hline
		$1/\alpha_h$ & 0.570 & N/A & N/A\\ 
		$1/\alpha_v$ & 1.420 & N/A & N/A\\ 
		$1/\gamma$ & 0.860 & N/A & N/A \\ \hline
		\multicolumn{4}{|l|}{Fixed Initial Conditions} \\ \hline
		$E_{h}(0)$ & 0 & N/A & N/A \\
		$U_{h}(0)$ & 1 & N/A & N/A \\
		$D_{h}(0)$ & 0 & N/A & N/A \\
		$R_{h}(0)$ & 0 & N/A & N/A \\
		$E_{v}(0)$ & 0 & N/A & N/A \\
		$I_{v}(0)$ & 1 & N/A & N/A \\ \hline
		\multicolumn{4}{|l|}{Active Parameters} \\ \hline
		$\beta$ &N/A& 1---10 & 4.745  \\
		$\delta$ &N/A&1---300& 275.979 \\
		$\mu$ & N/A & 1---6 & 2.712 \\ 
		$S_h(0)$ &N/A& $1,000,000$---$4,000,000$ & $3,980,441.000$ \\
		$S_v(0)$ & N/A& $1,000$---$5,000$& $3,789.000$ \\ \hline
		\multicolumn{4}{|c|}{\textbf{Chikungunya}} \\ \hline\hline
		\multicolumn{4}{|l|}{Fixed Parameters} \\ \hline
		$1/\mu$ & 0.369 & N/A & N/A \\ 
		$1/\alpha_h$ & 0.570 & N/A & N/A\\ 
		$1/\alpha_v$ & 1.420 &N/A& N/A\\ 
		$1/\gamma$ &0.860&N/A& N/A \\ \hline
		\multicolumn{4}{|l|}{Fixed Initial Conditions} \\ \hline
		$E_{h}(0)$ & 0 & N/A & N/A \\
		$U_{h}(0)$ & 1 & N/A & N/A \\
		$D_{h}(0)$ & 0 & N/A & N/A \\
		$R_{h}(0)$ & 0 & N/A & N/A \\
		$E_{v}(0)$ & 0 & N/A & N/A \\
		$I_{v}(0)$ & 1 & N/A & N/A \\ \hline
		\multicolumn{4}{|l|}{Active Parameters} \\ \hline
		$\beta$ &N/A& 1---10 & 4.537  \\
		$\delta$ &N/A&1---25&  20.000\\
		$S_h(0)$ &N/A& $1,000,000$---$2,000,000$ &   $1,740,058.000$ \\
		$S_v(0)$ & N/A&$1,000$---$50,000$& $1,167.000$ \\ \hline
	\end{tabular}
\end{table}
\begin{table}
	\centering
	\caption{Scores for information criteria: Akaike information criteria (AIC) and Bayesian information criteria (BIC). Model 1 has only one infectious compartment for hosts (not shown in text). Model 2 divides infectious hosts into undiagnosed and diagnosed compartments, 
	as defined by equations (\ref{sh})--(\ref{iv}).}
	\label{AICtable}
	\begin{tabular}{c|c|c|c}
	\hline
	{\bf Model} & {\bf Dataset} & {\bf AIC Score} & {\bf BIC Score} \\
	\hline
	Model 1 & Dengue & 1806.240 & 1814.045 \\
	Model 2 & Dengue & 1822.898 & 1832.655 \\
	\hline
	Model 1 & Chikungunya & 2175.164 & 2181.018 \\
	Model 2 & Chikungunya & 2163.279 & 2171.084 \\
	\hline
	\end{tabular}
\end{table}

\begin{figure}[h]
\centering
\includegraphics[height=2in,width=4.5in]{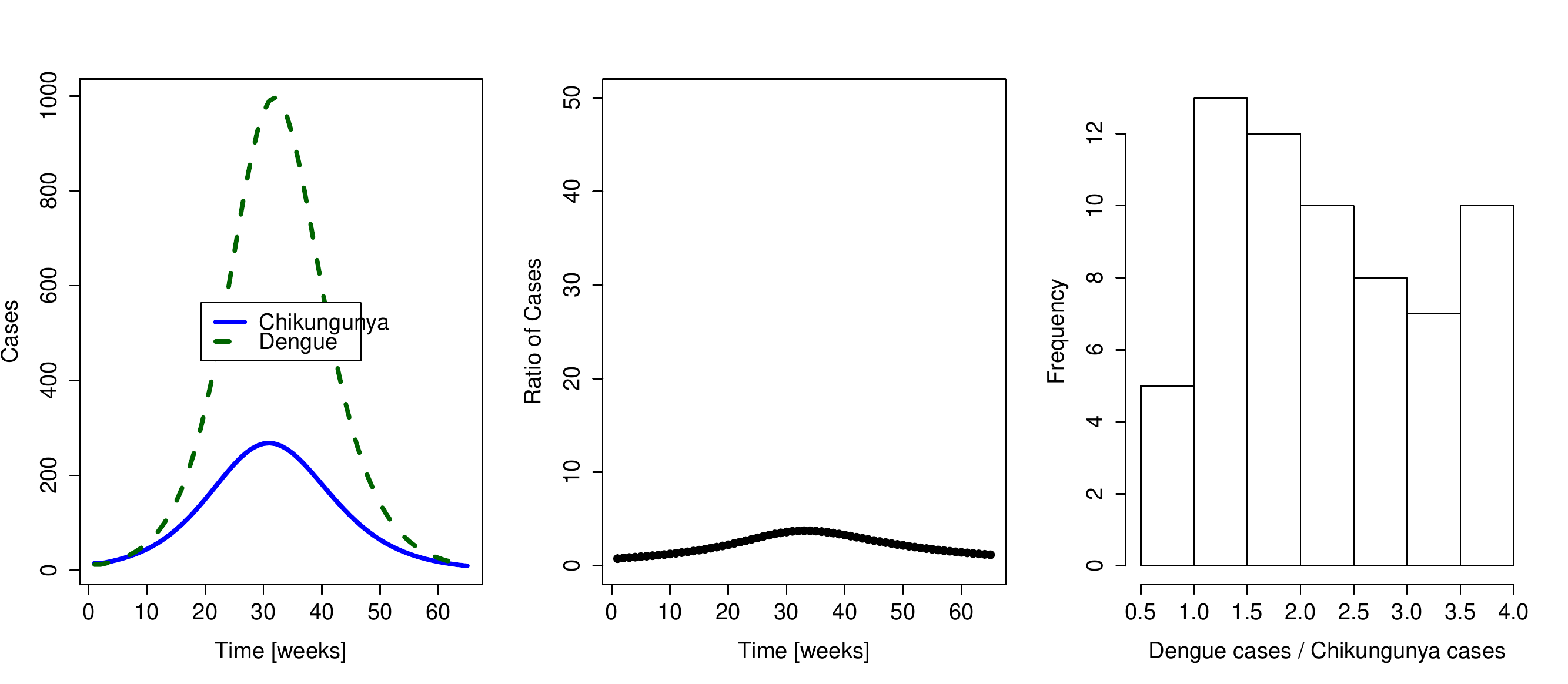}
\caption{Left panel: Model prediction for dengue (dashed) and chikungunya (solid) cases versus time. Middle panel: Time plot of the predicted dengue cases over chikungunya cases. Right panel: Histogram of the ratio between dengue and chikungunya predicted cases at each time point.}
\label{ratioDC}
\end{figure}


We also compare the model defined in equations (\ref{sh})--(\ref{iv}) with a particular model obtained by putting together the groups $D_h$ and $U_h$ in a single compartment $I_h$ (the equations
are not shown here). The statistical comparison was made through the AIC \cite{banks2014,akaike} and BIC \cite{burnham}
scores, both widely used in the literature of model comparison (see \cite{banks2014} and references therein). Both measures employ the assumption of Poisson-distributed data in this particular application, since the likelihood in their respective formulas is evaluated in the point estimates obtained in the optimization process (see Table~\ref{partable}). It is important to note that the BIC measure penalizes those models with a large number of parameters, in order to guarantee the choice of parsimonious alternatives.  A summary of the information
criteria scores is given in Table \ref{AICtable}.

Figure \ref{ratioDC} explores the question:  would it be possible that 
the number of cases of dengue scales as a multiple of the chikungunya cases?
The left panel displays the best fit plots over a time scale from week 0 to 60, since
this is a simulated scenarios the time window was chosen as such for better resolution.
The ratio of dengue divided by chikungunya cases, as predicted by best fit curves,
is depicted in the middle panel of Figure \ref{ratioDC}.  Each value of this ratio
is considered as a sample and such samples have a mean and median equal to
2.230 and 2.098, respectively (see right panel of Figure \ref{ratioDC}).  Therefore, 
this question is answered by saying that on average $z^{D}(t) = 2.230 z^{C}(t)$, i.e., chikungunya
cases $z^{C}(t)$ scaled up by a factor of 2.230 to match dengue cases $z^{D}(t)$.


\section{Discussion} \label{disc}
We used a single-outbreak deterministic model to estimate parameters using reported weekly cases for the dengue and chikungunya viruses from the 2015-2016 outbreak in Costa Rica. The reported data shows atypical behaviour where the dengue cases grows exponentially fast,  (see Figures \ref{data}--\ref{initial_epi_growth}), during the first 30 weeks of the season. This behaviour is unusual not due to the number of reported cases but rather that the chikungunya virus was present throughout the 2015 outbreak. This has been a challenge for public health officials mainly because throughout the last 20 years the vector {\it Ae. aegypti} was only transmitting dengue in the region. Once another disease transmitted by the same vector is introduced this makes the process of diagnosing patients very difficult. Most cases are clinically diagnosed since symptoms for both diseases are similar it is a challenge to determine which disease the individual may be infected with. Laboratory diagnosed cases are expensive and are the least of the reported cases.

Based on the point estimates obtained from fitting the model to incidence data, we can conclude that it is likely that dengue cases were misdiagnosed and chikungunya cases were under reported in the 2015-2016 outbreak. As the disease was introduced in the country and public health officials were unaware of the disease and the similarity of symptoms between the two viruses, this can possibly lead to a misdiagnosis.  

The information criteria scores that we computed in Table~\ref{AICtable} confirm that the compartments $U_h$ (infectious and undiagnosed) and $D_h$ (infectious and diagnosed) are well-suited for the dataset ‘Chikungunya’. This is the dataset with poor quality and anomalies, and in such case the need of an undiagnosed compartment pays off (better score). In a way these scores are reminiscent of the $R^2$ statistic in classical linear regression analysis. On the other hand, the dataset ‘Dengue’ has more quality (no expected anomalies) and in such case one compartment suffices to describe the data. In other words, there is no need to increase the complexity of the model.

In conclusion, our results show that misdiagnosis of cases can have a major impact on the long term dynamics of dengue and chikungunya. Our model highlights the importance of undiagnosed cases in a dengue/chikungunya outbreak as well as the importance of epidemiological surveillance and correct diagnosis. Routine detection of each virus in both vectors and hosts will be valuable in order to prevent major outbreaks and gauge the severity of the response that is required to combat any potential outbreak.  

\section*{Acknowledgements}
A. C.-A. thanks the support of the scholarship program {\em Preparation of Data Driven Mathematical Scientists for the Workforce}, housed by East Tennessee State University.


\begin{thebibliography}{22}
%

    \bibitem{akaike} Akaike, H. (1974). A new look at the statistical model identification. IEEE transactions on automatic control, 19(6), 716--723.

   \bibitem{banks2014} Banks, H. T., Hu, S., \& Thompson, W. C. (2014). Modeling and inverse problems in the presence of uncertainty. CRC Press.
    \bibitem{burnham} 
   Burnham, K. P., Anderson, D. R.  Model Selection and Multimodel Inference: A Practical Information-Theoretic Approach (2nd ed.), Springer-Verlag. (2002)
   
    \bibitem{byrd95} Byrd, R. H., Lu, P., Nocedal, J., \& Zhu, C. (1995). A limited memory algorithm for bound constrained optimization. 
    SIAM Journal on Scientific Computing, 16(5), 1190--1208.

    \bibitem{cdcchikv} Center for Disease Control, Chikungunya Virus. {\em Accessed January 2017.} \\
    \url{http://www.cdc.gov/chikungunya/}
 %
    \bibitem{cdcdengue} Center for Disease Control, Dengue.  {\em Accessed January 2017.}\\
    	\url{http://www.cdc.gov/dengue/}
	
    \bibitem{chahar2009} 
    Chahar, H. S., Bharaj, P., Dar, L., Guleria, R., Kabra, S. K., \& Broor, S. (2009). Co-infections with chikungunya virus and dengue virus in Delhi, India. 
    Emerg. Infect. Dis., 15(7), 1077--80.

    \bibitem{chastel2012} Chastel, C. (2012). Eventual role of asymptomatic cases of dengue for the introduction and spread of dengue 
    viruses in non-endemic regions. Frontiers in Physiology, 3, 70.    
    
    \bibitem{chowell2006b} Chowell, G., Ammon, C. E., Hengartner, N. W., \& Hyman, J. M. (2006). Transmission dynamics of the great influenza pandemic of 1918 in Geneva, Switzerland: assessing the effects of hypothetical interventions. Journal of theoretical biology, 241(2), 193--204.

    \bibitem{chowell2004} Chowell, G., Castillo-Chavez, C., Fenimore, P.W., Kribs-Zaleta, C.M., Arriola, L., Hyman, J.M., {Model Parameters and 
     Outbreak Control for SARS}. Emerging Infectious Diseases. {\bf 10}(7):1258--1263, 2004.
    
 
    \bibitem{chowell2006} Chowell, G., \& Sanchez, F. (2006). Climate-based descriptive models of dengue fever: the 2002 epidemic in Colima, Mexico. Journal of environmental health, 68(10), 40.    

    \bibitem{esteva1998} Esteva, L., \& Vargas, C. (1998). Analysis of a dengue disease transmission model. Mathematical biosciences, 150(2), 131-151.

%
%
    \bibitem{hapuarachchi2008} Hapaurachchi, H. A. C., Bandara, K. B. A. T., Hapugoda, M. D., Williams, S., \& Abeyewickreme, W. (2008). Laboratory 
    confirmation of dengue and chikungunya co-infection. Ceylon Medical Journal, 53(3).    
    
    
    \bibitem{harris2000} Harris, E., Videa, E., Pérez, L., Sandoval, E., Téllez, Y., Perez, M. L., ... \& Delgado, M. A. (2000). Clinical, epidemiologic, and virologic 
    features of dengue in the 1998 epidemic in Nicaragua. The American journal of tropical medicine and hygiene, 63(1), 5-11.
        

    \bibitem{Holland92} Holland, J. Adaptation in natural and artificial systems: an introductory analysis with applications to biology, 
    control and artificial intelligence. MIT Press. 1992. 
%
%
%
%
    \bibitem{manore2014} Manore, C. A., Hickmann, K. S., Xu, S., Wearing, H. J., \& Hyman, J. M. (2014). 
    Comparing dengue and chikungunya emergence and endemic transmission in A. aegypti and A. albopictus. Journal of theoretical biology, 356, 174--191.    

    \bibitem{mir2014} Mir, D., Romero, H., de Carvalho, L. M. F., \& Bello, G. (2014). Spatiotemporal dynamics of DENV-2 Asian-American 
    genotype lineages in the Americas. PloS one, 9(6), e98519.

    \bibitem{mscr} Costa Rica's Department of Health and Human Services
    (``Ministerio de Salud de Costa Rica"), health surveillance (``Vigilancia de la salud"), analysis of health status
    (``Analisis de situacion de salud").  {\it Accessed November, 2016}.\\
    \url{https://www.ministeriodesalud.go.cr}
   
    \bibitem{murillo2014} Murillo, D., Holechek, S. A., Murillo, A. L., Sanchez, F., \& Castillo-Chavez, C. (2014). Vertical transmission in a 
    two-strain model of dengue fever. Letters in Biomathematics, 1(2), 249--271.

%
    \bibitem{olivia2005historyofdengue} Dick, O. B., San Martín, J. L., Montoya, R. H., del Diego, J., Zambrano, B., \& Dayan, G. H. (2012). The history 
    of dengue outbreaks in the Americas. The American journal of tropical medicine and hygiene, 87(4), 584--593.    

%
\bibitem{pawelek} Pawelek, K. A., Niehaus, P., Salmeron, C., Hager, E. J., \& Hunt, G. J. (2014). Modeling dynamics of Culex pipiens complex 
populations and assessing abatement strategies for West Nile virus. PloS one, 9(9), e108452.


%
%
%
    \bibitem{sanchez2006} Sanchez, F., Engman, M., Harrington, L., \& 
    Castillo-Chavez, C. (2006). Models for dengue transmission and control. Contemporary Mathematics, 410, 311.    


    \bibitem{fsanchez2012} Sanchez, F., Murillo, D., \& Castillo-Chavez, C. (2012). Change in Host Behavior and its Impact on the Transmission 
    Dynamics of Dengue. In BIOMAT 2011 (pp. 191--203). World Scientific.  
    
    \bibitem{scrucca} Scrucca, L. (2013). GA: a package for genetic algorithms in R. Journal of Statistical Software, 53(4), 1--37.

%
%
%
%
%
%
%
\end{thebibliography}


\end{document}